\documentclass{aa}  

\usepackage{amssymb,amsmath}
\usepackage[version=3]{mhchem}
\usepackage{threeparttable}
\usepackage{chemformula}
\usepackage{hyperref}
\hypersetup{colorlinks,urlcolor=blue}
\usepackage{graphicx}
\usepackage{txfonts}

\begin{document} 

\title{A major asymmetric ice trap in a planet-forming disk}
\subtitle{III. First detection of dimethyl ether}

\author{Nashanty G.C. Brunken \inst{1}, Alice S. Booth \inst{1}, Margot Leemker \inst{1},  Pooneh Nazari \inst{1}, Nienke van der Marel \inst{1}, Ewine F. van Dishoeck \inst{1,2}}

\institute{Leiden Observatory, Leiden University, 2300 RA Leiden, the Netherlands \\
\email{abooth@strw.leidenuniv.nl} 
\and
Max-Planck-Institut f\"{u}r Extraterrestrishe Physik, Gie{\ss}enbachstrasse 1, 85748 Garching, Germany}

\titlerunning{COMs in the IRS~48 disk}
\authorrunning{N. G.C. Brunken et al.}

\date{Received; Accepted}

\abstract{The complex organic molecules (COMs) detected in star-forming regions are the precursors of the prebiotic molecules that can lead to the emergence of life. By studying COMs in more evolved protoplanetary  disks we can gain a better understanding of how they are incorporated into planets. This paper presents ALMA band 7 observations of the dust and ice trap in the protoplanetary disk around Oph IRS 48. We report the first detection of dimethyl ether (\ce{CH_3OCH_3}) in a planet-forming  disk and a tentative detection of methyl formate (\ce{CH_3OCHO}). We determined column densities for the detected molecules and upper limits on non-detected species using the CASSIS spectral analysis tool. The inferred column densities of \ce{CH_3OCH_3} and \ce{CH_3OCHO} with respect to methanol (\ce{CH_3OH}) are of order unity, indicating unusually high abundances of these species compared to other environments. Alternatively, the \ce{^{12}CH_3OH} emission is optically thick and beam diluted, implying a higher \ce{CH_3OH} column density and a smaller emitting area than originally thought. The presence of these complex molecules can be explained by thermal ice sublimation, where  the dust cavity edge is heated by irradiation and the full volatile ice content is observable in the gas phase. This work confirms the presence of oxygen-bearing molecules more complex than \ce{CH_3OH} in protoplanetary  disks for the first time. It also shows that it is indeed possible to trace the full interstellar journey of COMs across the different evolutionary stages of star, disk, and planet formation.}


\maketitle


\section{Introduction} 

Complex organic molecules (COMs) are the precursors of prebiotic molecules, and thus understanding their formation and evolution will help us gain more insight into how life originated in our own Solar System \citep{2012A&ARv..20...56C}. With new facilities such as the Atacama Large Millimeter/submillimeter Array (ALMA) and the Rosetta Orbiter Spectrometer for Ion and Neutral Analysis (ROSINA) on the Rosetta mission, we are now able to compare chemistry across a range of astronomical environments and get a better understanding of the chemical history of COMs throughout the entire star and planet-formation process, including comets \citep[e.g.][]{2019MNRAS.490...50D}. It is crucial to study COMs in planet-forming disks in order to understand how the material in the disk is incorporated into planets and what degree of complexity is present at the epoch of planet formation \citep{2021PhR...893....1O}.

The formation of most of these COMs is thought to occur in cold molecular clouds \citep{2015ARA&A..53..541B}. During this time, atoms and simple molecules such as CO will stick to the dust grains forming an ice layer and undergo chemical reactions \citep[e.g.][]{doi:10.1146/annurev-astro-082708-101654, 2018A&A...617A..87C, 2021NatAs...5..197I}. The products are subsequently released back into the gas phase if there is an increase in temperature resulting in thermal desorption. Additionally, molecules will also return to the gas phase via other processes such as UV photodesorption, but this can lead to the fragmentation of the molecule when it enters the gas phase \citep{garrod2006gas, 2016A&A...592A..68C}. COMs are therefore especially abundant in the gas phase in young warm systems where they are easily detected because of thermal sublimation ($T_{\mathrm{dust}}$ > 100~K) \citep[e.g.,][]{Bergner_2017, jorgensen2018alma, 2020A&A...639A..87V, 2021arXiv211107573M, 2020A&A...635A.198B}. This is in contrast to older protoplanetary disks which are colder and thus the COMs remain frozen on dust grains in the bulk of the disk and are therefore more often undetectable in the gas phase with ALMA \citep{2020arXiv200808106V}. However, COMs are expected to be abundant in protoplanetary disk ices and there is some evidence for this in the outbursting protostellar source V883 Ori which is rich in COMs \citep{van_t_Hoff_2018,lee2019ice}. 

The current situation is that, in protoplanetary disks of more than  1~Myr old,  even the most abundant COM, methanol (\ce{CH_3OH}), is difficult to detect. \ce{CH_3OH} is a cornerstone in the chemistry leading to many larger complex organic molecules \citep{oberg2009formation}. \citet{walsh2016first} presented the first detection of \ce{CH_3OH} in the TW Hya protoplanetary disk. However, the fractional abundance relative to \ce{H_{2}} is very low  (3 $\times$ \ce{10^{-12}} - 4 $\times$ \ce{10^{-11}}), indicating a chemical origin in the gas phase via inefficient and fragmenting non-thermal desorption of the ices rather than thermal sublimation \citep{walsh2017methanol}. \citet{carney2019upper} also provided an upper limit on the abundance of \ce{CH_3OH} in the Herbig Ae disk HD~163296 of $\textless$ 1.6 $\times$ \ce{10^{-12}} relative to \ce{H_{2}}.
For comparison, \ce{CH_3OH} abundances in hot protostellar cores are typically of order $10^{-6}$, comparable to those in ices \citep{2015ARA&A..53..541B}.
 
More recently, \citet{booth2021inherited} detected \ce{CH_3OH} for the first time in a warm Herbig transition disk. In comparison to the ringed \ce{CH_3OH} emission in TW~Hya the \ce{CH_3OH} in the HD~100546 disk originates from the inner 50~au of the disk and its likely origin is thermal desorption. Because Herbig Ae/Be sources like this one are inherently warm, which prevents freeze-out of the precursor CO, in situ formation of the \ce{CH_3OH} is unlikely. Instead the presence of \ce{CH_3OH} in the disk can be explained via the inheritance of COM-rich ices from colder parent molecular clouds. 

Also, another Herbig source was revealed to have a rich observable chemistry: the IRS 48 transition disk \citep{van2014warm, van2021major, booth2021major}.  What makes this disk particularly interesting is the fact that it contains a highly asymmetric dust trap of large grains ($\gtrsim$ 0.1 mm) on the southern side of the star, making it the most asymmetric disk detected to date \citep{van2013major, 2021AJ....161...33V}. \citet{van2021major} report the detection of \ce{CH_3OH} and formaldehyde (\ce{H_{2}CO}) in this disk. The emissions have the same crescent shape as the dust continuum, showing for the first time the direct link between a dust trap and COMs. This coincidence was hinted at with low signal-to-noise \ce{H_2CO} observations \citep{van2014warm} but is now confirmed.
The bulk of the ice reservoir of the IRS48 disk  is constrained to the larger dust grains, and because of UV irradiation from the central star, the dust temperature increases enough to liberate the \ce{CH_3OH} from the ices. \citet{booth2021major} additionally report the detection of \ce{SO_2} in the IRS~48 dust trap, the first detection of this molecule in a protoplanetary disk, along with detection of \ce{SO}. The detection of these molecules supports the presence of oxygen-rich gas where the C/O $<$ 1 because of sublimated ices.

In this paper, we report the analysis of ALMA data of IRS 48 including the first detection of dimethyl ether (\ce{CH_{3}OCH_{3}}) in a protoplanetary disk and a tentative detection of methyl formate (\ce{CH_3OCHO}). \ce{CH_3OCH_3} is the largest complex organic molecule that has been detected in a protoplanetary disk to date. 
We also report the first detection of nitric oxide (\ce{NO}) in 
a protoplanetary disk, which will be analysed in a future paper. 
Our paper is structured as follows: In Section \ref{sec:2} we describe our observational methods and in Section \ref{sec:3} we show our data analysis and provide the values for the derived column densities. In Section \ref{sec:4} we discuss the chemistry of the detected species, compare abundances to other astronomical environments, and determine upper limits for other molecules covered in the data. Finally, in Section \ref{sec:5} we give a short summary and provide conclusions.

\section{Observations}
\label{sec:2}

Our data were taken with ALMA. The Band 7 line data ($\sim$ 0.8mm) were taken on August \ce{18,^{}} 2018 (2017.1. 00834.S, PI: Adriana Pohl), and the continuum data presented in Figure~1 were taken in June and August 2015 (2013.1.00100.S, PI: Nienke van der Marel). \citet{ohashi2020solving} provide a full description of the line data calibration. In papers I and II, we cover the detections of \ce{CH_{3}OH}, \ce{H_{2}CO}, \ce{SO}, and \ce{SO_{2}} \citep{van2021major, booth2021major} and in this paper we present the detection of \ce{CH_{3}OCH_{3}} and NO, and investigate other tentative detections and upper limits.

Data reduction was done using the Common Astronomy Software Applications (CASA)\footnote[1]{\url{https://casa.nrao.edu/index.shtml}} version 5.7.0. The spectral windows have channel widths of $\sim$ 1.7 \, km$~\mathrm{s^{-1}}$ and a beam size of  0\farcs55$\times$0\farcs42 (PA = 80$^{\circ}$).
 The spectral windows have central frequencies of 349.7, 351.5, 361.6, and 363.5 GHz, respectively, with SPW1 from 349.79 to 350.66 GHz, SPW2 from 350.60 to 352.47 GHz, SPW3 from 360.68 to 362.55 GHz, and SPW4 from 362.61 to 364.47 GHz. We imaged the data with the \texttt{tclean} function in CASA using a Briggs weighting with a robust value of 0.5. 
 The image was recentred to the star position using the phase centre parameter in CASA and was set to ICRS~16:27:37.17~-24:30:35.55. We used a Keplerian mask over the region of emission at a distance of 136~pc \citep{2021A&A...649A...1G}, 
 an inclination angle of \ce{50^{$\circ$}},  and a position angle of \ce{100^{$\circ$}} \citep{2021AJ....161...33V}.

The cleaned images were subsequently stacked using GoFish version 1.3.6 \citep{teague2019gofish} in order to increase the signal-to-noise ratio. This method makes it possible to identify potential weak lines and also distinguish between lines that are blended (very close in frequency). We extract spectra over the whole azimuth of the disk although the lines are co-spatial with the dust trap. This is done because the observations are not well spatially resolved. 
The spectra for the four spectral windows are shown in the Appendix Figures~B1-B4. 

\section{Analysis}
\label{sec:3}

\subsection{Spectral analysis}

\begin{figure*}[h!]
    \centering
     \includegraphics[width=0.9\hsize]{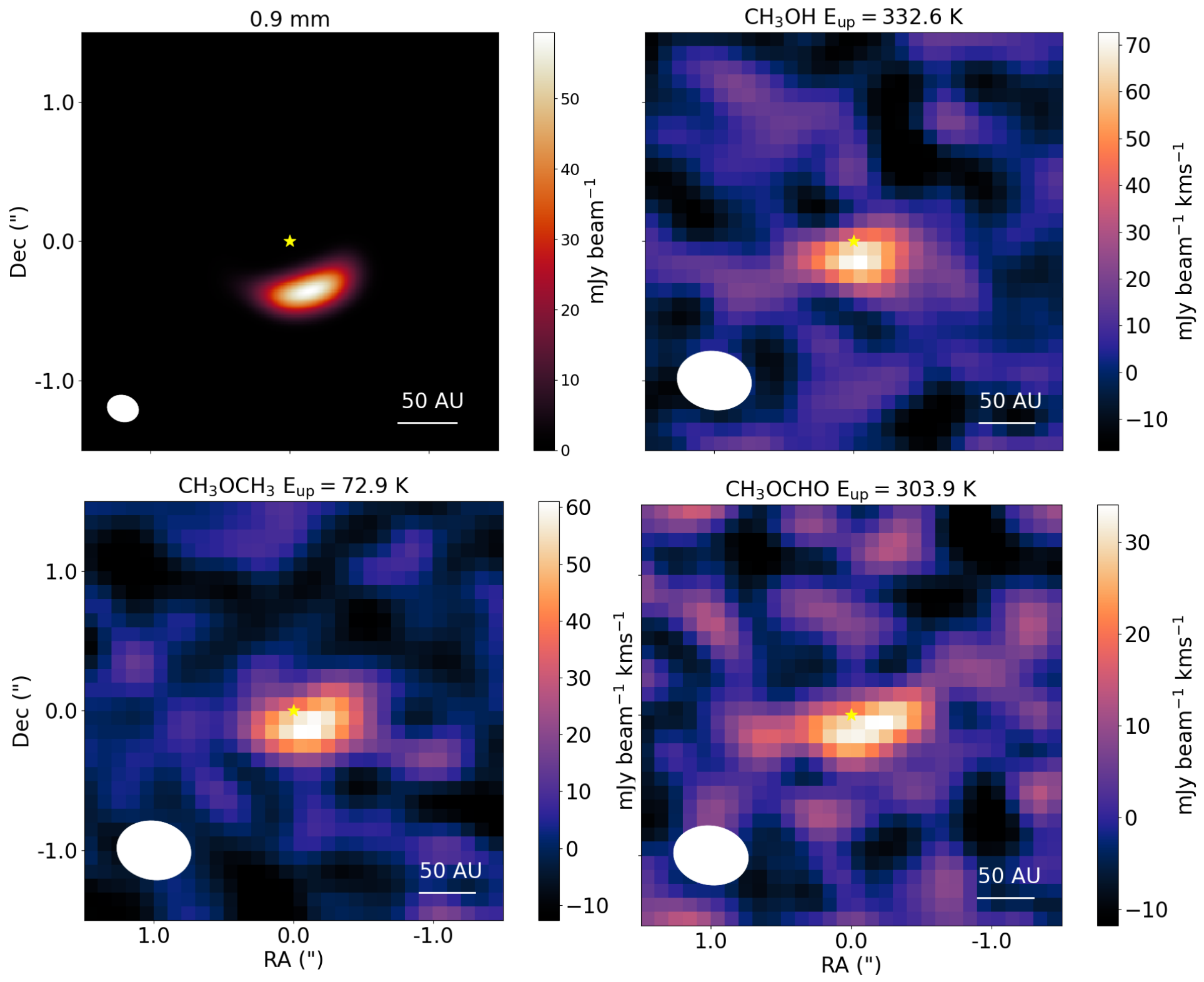}
    \caption{Integrated intensity maps of the 0.9 mm continuum emission and a subset of the detected molecular lines listed in Table \ref{Tab:1}. Top right is the \ce{CH_3OH} \ce{16_{1,15}} - \ce{16_{0,16}}, bottom left is the \ce{CH_3OCH_3} \ce{11_{3,8}-10_{2,9}}, and bottom right is the \ce{CH_3OCHO} \ce{32_{2,30}-32_{2,29}} and \ce{32_{3,30}-32_{3,29}} blend. The beam is shown in the bottom left corner and a scale bar is shown in the bottom right corner. } 
    \label{fig:mom0maps}
\end{figure*}

The stacked, continuum-subtracted spectra were analysed using the CASSIS\footnote[2]{\url{http://cassis.irap.omp.eu}} spectral analysis tool version 5.1.1 \citep{vastel2015cassis} in a similar way to that used by \citet[e.g.,][]{nazari2021complex}. The flux densities were first converted to brightness temperature units and local thermodynamic equilibrium (LTE) conditions were assumed in order to derive the column densities and excitation temperatures. We made use of the Cologne Database for Molecular Spectroscopy (CDMS) \citep{muller2001cologne, muller2005cologne} and the Jet Propulsion Laboratory (JPL) database \citep{pickett1998submillimeter} for molecular information. 
In Table \ref{Tab:1} we list the transitions of the detected species. The integrated intensity maps of selected lines are presented in Figure~\ref{fig:mom0maps}.
The search for other potential features in the spectra was carried out by making a selection of commonly detected COMs in other environments and only taking into account lines with ${E_{\mathrm{up}}}$ $\leq$ 400 K and $A_{\mathrm{ul}}$ $\geq$ \ce{1 \times 10^{-6}} \ce{s^{-1}}. For this selection of molecules, we modelled the spectra in CASSIS by assuming an excitation temperature of \ce{100}~K motivated by the rotational temperature derived by \citet{van2021major} for the \ce{CH_3OH}. We also calculated the best-fit column density at both 70~K and 250~K to have an estimate of the column density error for the detected species, which is typically a factor of two. The absolute calibration error is much smaller, of order 10\%, and this will cancel out in abundance ratios. We used a FWHM of $\sim$ 7 km~s$^{-1}$ based on the line width of a strong \ce{CH_3OH} line and a source size of 1.4 $\times$ \ce{10^{-11}} sr based on the 5 $\sigma$ emission continuum of the disk (Figure~1), the same as in \citet{van2021major, booth2021major}. Using these variables we derived column densities and upper limits. 
We note that the inferred column density, $N$,  scales inversely with the assumed emitting area, $\Omega_{\mathrm{source}}$  \citep{1999ApJ...517..209G}, that is,  
\begin{equation}
N \propto \frac{1}{\Omega_{\mathrm{source}}}.
\end{equation}
In the case where the source does not fill the beam, the column density will be underestimated by a dilution factor:
\begin{equation}
\text {Dilutionfactor }^{-1}=\frac{\Omega_{\mathrm{source}}^{2}}{\Omega_{\mathrm{beam}}^{2}+\Omega_{\mathrm{source}}^{2}},
\end{equation}
where $\Omega_{\mathrm{beam}}$ is the beam size \citep[e.g.][]{2020A&A...639A..87V}.

\subsection{Revising the \ce{CH_3OH} column density}

We first modelled the \ce{CH_3OH} lines in our spectra based on the parameters derived in \citet{van2021major}. Using CASSIS we find models consistent with the data using a column density for \ce{CH_3OH} of 5 \ce{ \times 10^{14} \, cm^{-2}} and an excitation temperature of 100~K which is in agreement with the rotational diagram analysis of \citet{van2021major}. These results are shown in Figure~2. However, in this work, additional \ce{CH_3OH} transitions are detected and these are highlighted in Table~A1. Figure \ref{fig:mom0maps} shows the integrated intensity maps of a \ce{CH_3OH} line with an upper energy level of 333~K. We also detect two weaker lines in the stacked spectra that are better fit at a higher column density of 2~$\times$~\ce{10^{15}}\, cm$^{-2}$ and still at a temperature of 100~K (see Figure~2 for a comparison of the two models). These two lines are the \ce{9_{-5,4}} - \ce{9_{-4,6}} and \ce{3_{1,2}} - \ce{4_{2,2}} at 351.236~GHz ($E_{\mathrm{up}} =~$241~K, 
$\mathrm{log_{10}}(E_{\mathrm{A}}) = {-4.44}$~s$^{-1}$) and 361.236~GHz ($E_{\mathrm{up}} =~$339~K, $\mathrm{log_{10}}(E_{\mathrm{A}}) ={-3.58}$~s$^{-1}$) respectively.
Both lines were visible in our stacked spectra at the 2.5-3$\sigma$ level but neither one was detected in the channel maps above the 3~$\sigma$ level. These weak lines are reproduced at a different column density, which likely indicates that the emission traced by the stronger lines is optically thick. \citet{van2021major} calculate the optical depth of these lines and show they are optically thin for the assumed emitting area. The difference between this and our result can be resolved if the lines are optically thick and beam diluted, because the column density is inversely proportional to the assumed emitting area.
We also derived a 3$\sigma$ upper limit for the column density of \ce{^{13}CH_{3}OH} of $<$~5.5~$\times 10^{14}$~cm$^{-2}$. This gives a strict upper limit on the 
\ce{CH_{3}OH} column density of $\approx$~3.3~$\times 10^{16}$~cm$^{-2}$ assuming a \ce{^{12}C}/\ce{^{13}C} ratio of 60. 
This upper-limit is consistent with the column density found via the weakest \ce{CH_3OH} lines.
We use N(\ce{CH_3OH}) = $2\times10^{15}$ cm$^{-2}$ as a reference for comparisons. 

\subsection{Detection of dimethyl ether and methyl formate}

We detect two sets of blended lines for \ce{CH_3OCH_3}. See Table \ref{Tab:1} for the transition information,  Figure \ref{fig:mom0maps} for an intensity-integrated map of one of the sets of blended lines, and Figure \ref{fig:channelether} for the channel maps of both. We derive a column density of 1.5 \ce{ \times 10^{15} \, cm^{-2}} at an assumed excitation temperature of 100 K; see Figure \ref{fig:blowups} for the CASSIS model fits.  
The excitation temperature of \ce{CH_3OCH_3} may be lower than that of \ce{CH_3OH} \citep[e.g.][]{jorgensen2018alma} but given that only a few transitions are detected, we calculate the column densities over a range of excitation temperatures from 80 to 250~K (listed in Table~1).
We also found a tentative detection of \ce{CH_{3}OCHO} after modelling several features in our spectra. One such emission feature with a S/N above 3$\sigma$ noise level can be seen at 363.48 and 363.49 GHz (Figure \ref{fig:blowups}). 
We show a clear $>5$~$\sigma$ detection of a line in the channel maps at this frequency range (Figure~\ref{fig:channel maps}) and the integrated intensity map is shown in Figure~\ref{fig:mom0maps}. 
While the modelled spectrum of \ce{CH_3OCHO} does not provide an exact fit for the emission feature, it is the closest fit  found after considering other possible candidates.  
Furthermore, the model spectrum for \ce{CH_3OCHO} fits several other smaller features in the spectra (see Figures~\ref{fig:spw0},\ref{fig:spw1},\ref{fig:spw2},\ref{fig:spw3}). From these models we derive a best-fit column density of 1.3 $\times$ \ce{10^{15}} \ce{cm^{-2}} at an excitation temperature of 100 K.

\subsection{Other line detections and upper limits}

\begin{figure*}
      \vspace{0.5cm}
     \centering
     \includegraphics[width=0.85\hsize]{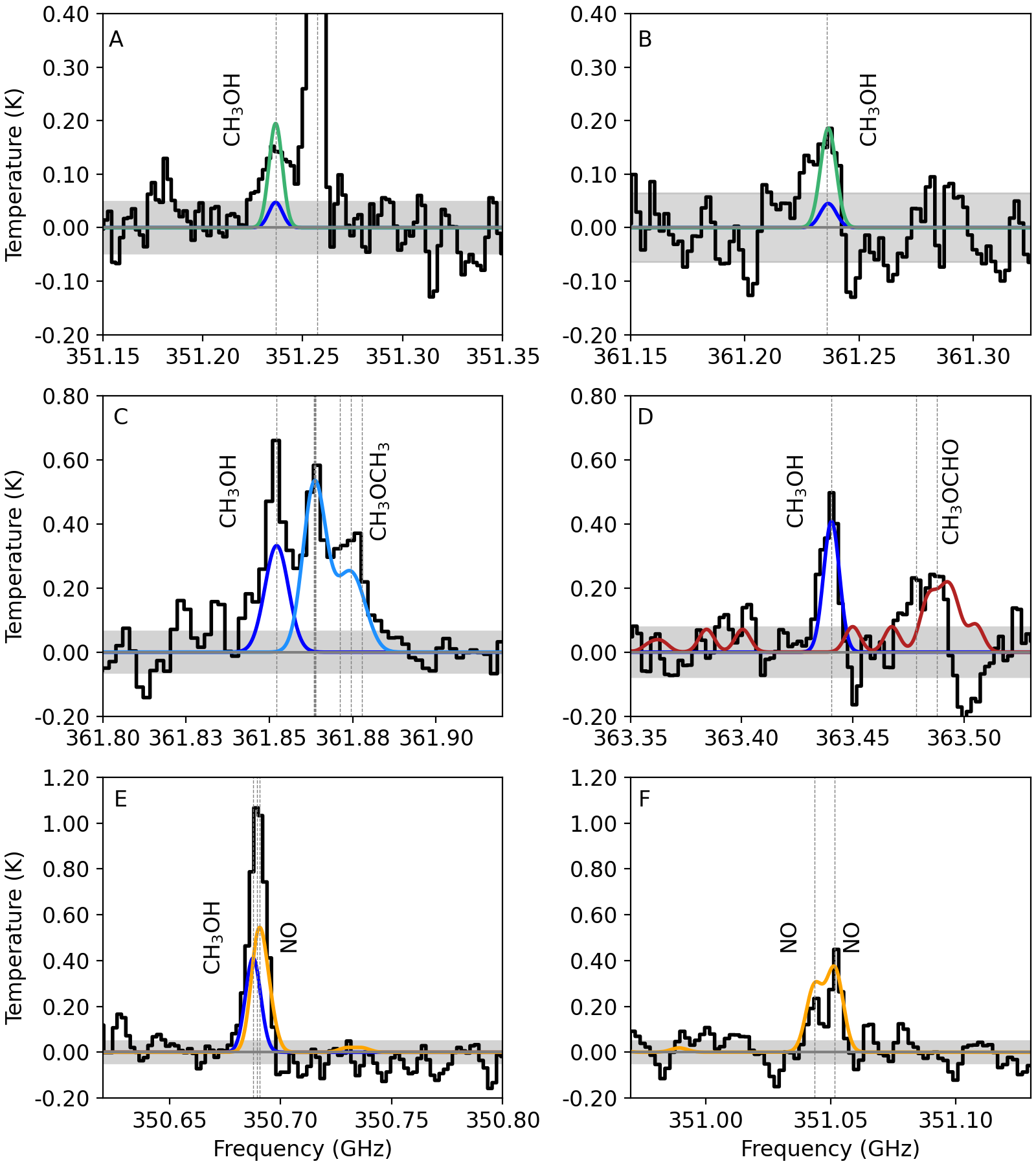}
     \caption{Stacked continuum-subtracted spectra (black lines) and CASSIS models (coloured lines) for the molecules detected in this work. Dashed lines mark the frequencies of the transitions as listed in Table~\ref{Tab:2} and the grey bar marks the $\pm 1 \sigma$ error calculated from the line-free channels in each spectral window. Panels A and B show the two weak \ce{CH_3OH} lines \ce{9_{-5,4}} - \ce{9_{-4,6}} and \ce{3_{1,2}} - \ce{4_{2,2}} with the 100~K CASSIS models at 5~$\times$~\ce{10^{14}} cm$^{-1}$ (blue) and 2~$\times$~\ce{10^{15}} cm$^{-1}$ (green). Panels C and D show  the best-fit models for the \ce{CH_3OCH_3} \ce{20_{1,20}-19_{0,19}} and \ce{11_{3,8}}-\ce{10_{2,9}} transitions and the \ce{CH_3OCHO} \ce{32_{3,30}}-\ce{31_{3,29}} and \ce{32_{3,30}}-\ce{31_{2,29}} transitions. In panel D the negative dip in the spectrum at $\approx$363.5~GHz may be an atmospheric absorption feature (\url{https://almascience.eso.org/about-alma/atmosphere-model}). Panels E and F show the best-fit models for NO covering the \ce{4_1}-\ce{4_3} and \ce{4_1}-\ce{4_4} transitions. Panels C, D, and E also show the \ce{CH_3OH} model for the strong lines.
}
     \label{fig:blowups}
 \end{figure*}

We detect a total of five transition lines for nitric oxide (Table \ref{Tab:1}). This is the first detection of NO in a protoplanetary disk. We were alerted to the possible presence of NO in our disk when we encountered a difficulty in fitting the bright \ce{CH_3OH} line at ~350.68 GHz (Figure \ref{fig:blowups}).  NO has two transitions at this frequency, but a single line was not enough to confirm a definitive detection of the molecule because this line is also blended with \ce{CH_3OH}. We are able to prove the presence of NO in the disk after successfully fitting an additional double feature at 351.04 and 351.05 GHz (Figure \ref{fig:blowups}). From the CASSIS spectral analysis models, we derive a best-fit total column density for NO of \ce{3 \times 10^{15}\, cm^{-2}} at an excitation temperature of 40~K. The NO lines have a low excitation temperature (36~K) compared to many of the COM lines detected and the lower temperature best fits the multiple lines. The NO lines will be more quantitatively analysed in a future paper.  

We also detect an additional \ce{SO_2} line at 363.16 GHz (Figure \ref{fig:spw0}) that was not reported in \citet{booth2021major}. The column densities from our spectral analysis are in agreement with their value.

Finally, we also derive upper limits for species that remain undetected in the IRS 48 disk but that have been observed in younger sources and other older disks such as formic acid (t-HCOOH), acetaldehyde (\ce{CH_{3}CHO}), and methyl cyanide (\ce{CH_{3}CN}) \citep[e.g.,][]{Bergner_2017, 2018ApJ...862L...2F,  2020A&A...639A..87V, 2021ApJS..257....9I}. These upper limits are listed in Table~1.

\section{Discussion}
\label{sec:4}

We find a wealth of molecular complexity in the IRS~48 disk, including the first detections of multiple molecules in disks. 
In this section, we discuss the chemical origin of the COMs, compare relative abundances to other environments, and consider the prospects for further complexity in the disk.

\subsection{Chemical origin of the COMs}
\label{section:4.4}

The observed \ce{CH_{3}OH} emission in the IRS~48 disk first presented by \citet{van2021major} is azimuthally co-spatial with the dust trap and peaking at slightly smaller radius. \citet{van2021major} proposed that the presence of \ce{CH_{3}OH} in the disk is due to thermal ice sublimation and that the ice reservoir is constrained to the larger millimetre-sized grains. Vertical mixing in the vortex may also help in lifting icy dust grains to the warm surface.  
\ce{CH_{3}OH} forms on CO ice via a sequence of H-addition reactions with key intermediates \ce{HCO} and \ce{H2CO} \citep{2009A&A...505..629F,2017MNRAS.467.2552C}.
Because the grain surface chemistry of \ce{CH_{3}OH} is related to both \ce{CH_3OCH_3} and \ce{CH3OCHO},  particularly in the presence of UV radiation,
we expect that both of these COMs also originate from the sublimating ices \citep{oberg2009formation},
\citet{garrod2006formation} and \citet{garrod2008complex} provide a theoretical model in which complex organic molecules, including \ce{CH_3OCH_3} and \ce{CH3OCHO}, can form via cold grain-surface reactions ($\leq$ 50 K) involving radicals:
\begin{equation}
    \ch{CH3O + HCO -> CH3OCHO } 
,\end{equation}
\begin{equation}
    \ch{CH3O + CH3 -> CH3OCH3} 
.\end{equation}
This model shows a common formation route from the methoxy precursor \ce{CH_3O}.
These pathways have also been shown to be present in laboratory
experiments \citep{2016MNRAS.455.1702C}.
These COMs could nevertheless be further enhanced due to UV irradiation of the ices from the central star resulting in  photodissociation of \ce{CH_3OH} \citep{oberg2009formation, walsh2014complex}, 
\begin{equation}
    \ch{CH3OH + h\ce{\nu} -> CH3 + OH ~~or~~CH3O + H}, 
\end{equation}
where the dissociation products can then recombine via reactions (1) and (2). 

\begin{figure}
     \includegraphics[width=\hsize]{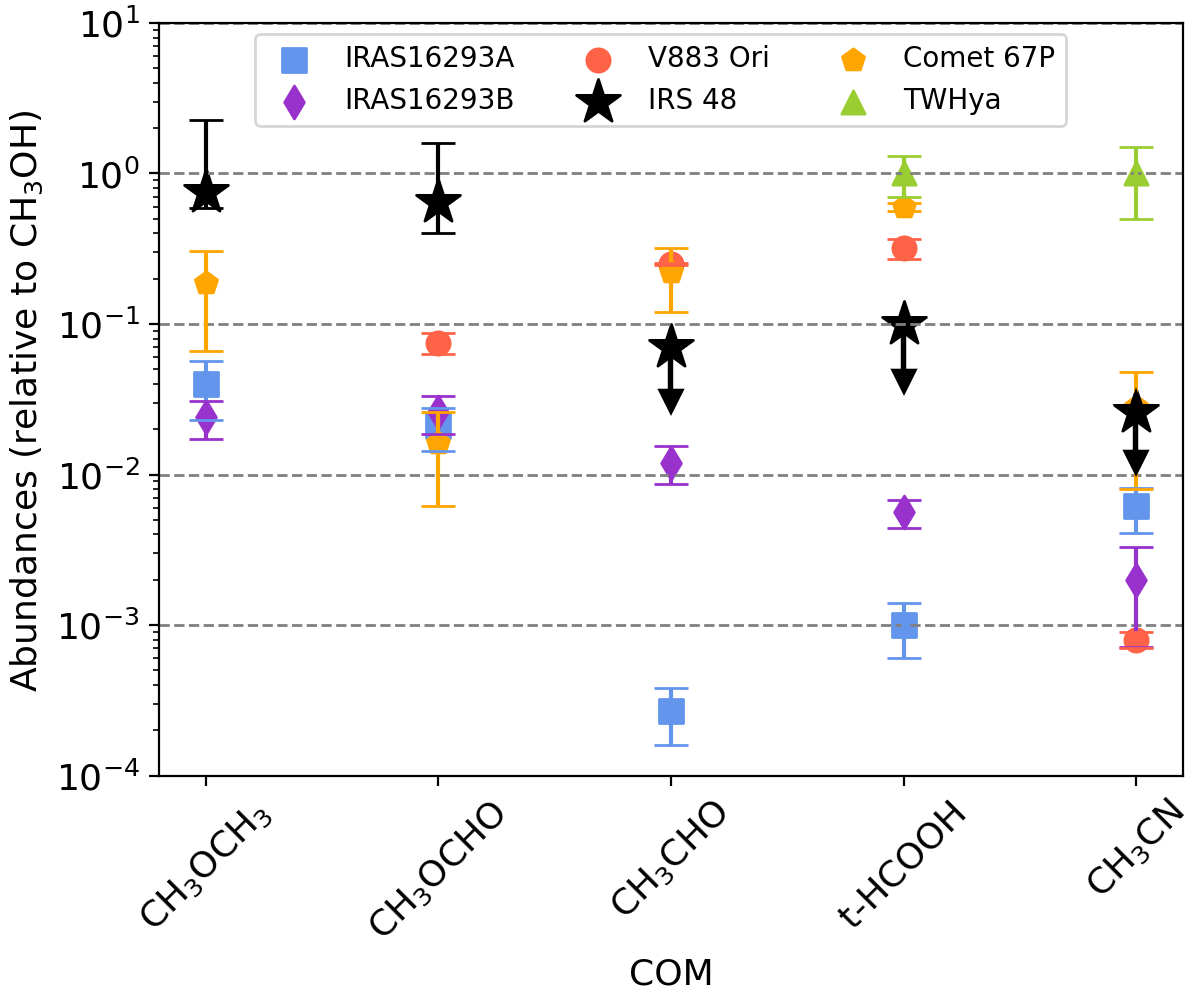}
     \caption{Abundances of commonly detected COMs relative to \ce{CH_3OH}. Solid squares show the detections in IRS~48 and arrows show the non-detected species for which upper limits on the column density are derived. The error bars on the IRS~48 points come from calculating the column densities over a range of excitation temperatures. For the other sources, see references in Section 4.} 
     \label{fig:barplots}
\end{figure}

\subsection{Comparisons to other environments}

\ce{CH_3OCH_3}  is the largest complex organic molecule detected in a protoplanetary disk. It has been detected in several other, younger sources \citep[e.g.][]{Taquet_2015, soma2018complex, bergner2018survey}. We compare our results as summarised in Table~1 (100~K column) and Figure~3 with the observed abundances in other astronomical environments including the class 0 protostellar binary IRAS 16293 A and B \citep{jorgensen2018alma,manigand2020alma}, the outbursting source V883 Ori \citep{lee2019ice}, and the comet 67P \citep{drozdovskaya2019ingredients}. 
The high observed \ce{CH_{3}OCH_{3}}/\ce{CH_{3}OH} ratio in the IRS~48 disk, of order unity, is different from that in the other sources by a factor of 5-10 (Figure \ref{fig:barplots}). The abundance derived for \ce{CH_3OCHO} also shows a similar trend to \ce{CH_3OCH_3}  in that \ce{CH_3OCHO} seems to be more abundant compared to the other sources. The high derived column-density ratios of \ce{CH_{3}OCH_{3}}/\ce{CH_{3}OH} and \ce{CH_{3}OCHO}/\ce{CH_3OH} in IRS~48 (Figure \ref{fig:barplots}) compared to other environments may be due to optically thick \ce{CH_{3}OH} emission that is beam diluted resulting in  over-estimated abundance ratios. 
To increase the optical depth in \ce{CH_3OH} to the amount that would make the ratio consistent with other sources, an area of order $10^{-12}$ sr would be needed. If the emission is constrained to the inner edge of the dust cavity, this 
would require a crescent shape for the emitting area with a length
of $\approx$1\farcs0,  corresponding to a width of $\approx$0\farcs1 in COM emission,
which could be resolved in future higher resolution data.
Detections of \ce{CH_3OH} isotopologues are needed to determine whether optical depth is indeed the cause of the difference or chemical processing in the disk relative to ices in dark clouds and young stars could be responsible for the enhanced chemical complexity in the UV-irradiated ice trap. 

The \ce{CH_{3}OCH_{3}}/\ce{CH_{3}OCHO} ratio in our disk is approximately 0.9 and this is consistent with what is observed in other sources across a full range of environments from star and disk formation to comets \citep{coletta2020evolutionary}. This adds further evidence that these two species are likely chemically related to one another and points towards ice formation of both molecules and therefore the inheritance of ices in the IRS~48 disk.

\subsection{Upper limits}

We also derived upper limits for the column densities of COMs that were previously detected in other sources. These molecules are listed in Table \ref{Tab:2}; see also Figure \ref{fig:barplots}. Although \ce{CH_{3}CHO}, t-HCOOH, and \ce{CH_3CN} have been detected in several sources, we note that they remain absent in our disk despite having formation routes via grain-surface chemistry \citep{walsh2014complex}. 

In particular, \ce{HCOOH} can form via the \ce{HCO} or \ce{HOCO} radicals and \ce{CH_3CHO} via the \ce{CH_3} and \ce{HCO} radicals. 
The non-detection of formic acid is potentially interesting, as in disk chemical models it is predicted to have 
a similar fractional abundance to \ce{CH_3OH} in the gas phase \citep{walsh2014complex}.
From the upper limit, we constrain this ratio to <10~\%. However, 
\citet{walsh2014complex} predict that the \ce{CH_3OH} ice column density is approximately ten times higher than \ce{HCOOH}. This could explain the non-detection of \ce{HCOOH} in our data if the sublimating ice reservoir is the primary origin of both molecules. TW Hya also has a detection of formic acid \citep{2018ApJ...862L...2F} and in this disk the \ce{t-HCOOH}/\ce{CH_3OH} abundance ratio is approximately unity. This is at least an order of magnitude higher than the 1\%-10\% seen in young stars and comets \citep[e.g.][]{drozdovskaya2019ingredients}. In comparison to IRS~48, where the observable chemistry appears to be dominated by ice sublimation, TW~Hya is a cold disk where small amounts of COMs in the gas phase are due to non-thermal desorption and/or gas-phase chemistry.

Similarly, with the detection of \ce{CH_{3}OCH3} we might also expect to have detected \ce{CH_{3}CHO}. The abundance of \ce{CH_{3}CHO} with respect to \ce{CH_3OH} is found to be about ten times lower thanthat of \ce{CH_{3}OCH3}. This difference is consistent with the results of \citet{2020A&A...639A..87V} who find lower abundances of \ce{CH_{3}CHO} compared to \ce{CH_{3}OCH3} in young protostellar envelopes.

Another molecule that is particularly interesting to look at is \ce{CH_{3}CN} as it has been detected in multiple protoplanetary disks \citep{loomis2018distribution,bergner2018survey,2021ApJS..257....9I}. The formation of \ce{CH_{3}CN} seems to be dominated by gas-phase chemistry but grain surface processes cannot be neglected \citep{loomis2018distribution}. In particular, gas-phase \ce{CH_3CN} is enhanced in environments with high C/O ratio. In TW Hya, the only disk with detections of both \ce{CH_3OH} and \ce{CH_3CN,} the \ce{CH_3CN}/\ce{CH_3OH} column density ratio is approximately unity. 
Unlike what we expect for IRS~48, in TW~Hya the observable \ce{CH_3CN} and \ce{CH_3OH} likely do not have the same chemical origin.
The \ce{CH_3CN} is primarily formed via gas-phase routes whereas the \ce{CH_3OH} most likely originates from the ices \citep{loomis2018distribution,walsh2016first, walsh2017methanol}.
In IRS~48 we have an upper limit of $\approx$10\%. 
This is in better agreement with the 1\%-10\% seen in comets and young stars \citep[see Figure 3 and e.g. ][]{Bergner_2017}. 

We also obtained upper limits on the deuterated form of methanol \ce{CH_2DOH}. The upper limit on this column density at an excitation temperature of 100~K is \ce{6.0 \times 10^{14}} cm$^{-2}$. This then gives an upper limit on the D/H of 10\% and this is consistent with the ratios seen in protostellar cores, young low-mass stars, and comet 67P ($\approx$1\%-10\%, e.g. \citealt{2020A&A...639A..87V, 2021MNRAS.500.4901D}).

\begin{table*}
\begin{center}
\caption{Derived column densities and upper limits}
\small
\begin{tabular}{clccccccccc}
\hline \hline
Species & Name & Column density ($\mathrm{cm^{-2}}$) &Column density ($\mathrm{cm^{-2}}$) \\
 &  & $T\mathrm{_{ex}}$ = 100~K & $T\mathrm{_{ex}}$ = 70 - 250~K \\

\hline
\ce{CH_{3}OH} & Methanol &  2~$\times 10^{15}$&{1.0 - 2.5} $\times 10^{15}$ \\
\ce{CH_{3}OCH_3} & Dimethyl ether & 1.5~$\times 10^{15}$  &1.5 - 3.5~$\times 10^{15}$ \\
\ce{CH_3OCHO} & Methyl formate &  1.3~$\times 10^{15}$ &1.0 - 2.5~$\times 10^{15}$ \\
\hline
\ce{^{13}CH_{3}OH} & Methanol &  < \ce{5.5 \times 10^{14}}& -  \\
\ce{CH_{2}DOH} & Methanol &  < \ce{6.0 \times 10^{14}}  & - \\
\ce{t-HCOOH} & Formic Acid & < \ce{2.0 \times 10^{14}}& - \\
\ce{CH_{3}CN} & Methyl cyanide & < \ce{1.3 \times 10^{13}} &  - \\
\ce{CH_{3}CHO} & Acetaldehyde  & < \ce{1.4 \times 10^{14}}  & -  \\
\ce{CH_{3}COOH} & Acetic acid &< \ce{3.4 \times 10^{15}}  & - \\
\ce{CH_{2}(OH)CHO} & Glycoaldehyde & < \ce{1.8 \times 10^{14}}&  - \\
\ce{CH_{3}SH} & Methyl mercaptan & < \ce{4.0 \times 10^{14}}&  - \\
\ce{D_{2}CO} & Formaldehyde & < \ce{4.5 \times 10^{13}}   &  - \\
\ce{HC_{3}N} & Cyanoacetylene & < \ce{2.2 \times 10^{13}} &  -  \\
\ce{HNC} & Hydrogen cyanide & < \ce{1.5 \times 10^{12}}&  -   \\
\ce{HNCO} & Isocyanic acid & < \ce{4.5 \times 10^{13}} &  -  \\
\ce{NH_2CH_2COOH} & Glycine, conf. I &  <\ce{4.5 \times 10^{15}} &  -  \\
\ce{NH_2CH_2COOH} & Glycine, conf. II & < \ce{5.5 \times 10^{14}} &  -   \\
\hline
\end{tabular}
\label{Tab:2}
\begin{tablenotes}\footnotesize
\item{Molecular information was obtained using the JPL and CDMS database \citep{muller2001cologne, muller2005cologne,pickett1998submillimeter}. The spectra were modelled using the CASSIS spectral analysis tool. For the detected lines, a range of $T_{\mathrm{ex}}$ from 70~K to 250~K was modelled. The 3 $\sigma$ upper limits were derived assuming a $T_{\mathrm{ex}}$ of 100~K. }
\end{tablenotes}
\end{center}
\end{table*} 

\subsection{Prospects for further complexity in the IRS 48 ice trap}

Confirmation of the \ce{CH_3OCHO} detection is needed because our models do not fit the emission feature very well and we only have one significant feature to fit given the frequency coverage of the observations. There are several other COMs that remain undetected in our disk and which should be the focus of future work. The molecules listed in Table \ref{Tab:2} are examples of species that should be searched for in future studies due to the fact that most have been detected in multiple protostellar sources. The detection of \ce{CH_3OCH_3}  and \ce{CH_3OCHO} alongside \ce{CH_3OH} implies a rich ice chemistry in the IRS~48 dust trap. Other COMs which have related formation routes via the radials \ce{HCO}, \ce{CH3O,} and \ce{CH2OH,} including ethylene-glycol, acetaldehyde, ethanol, and glycolaldehyde, should be subject of follow-up observations. 

\section{Conclusions}
\label{sec:5}

We analyzed ALMA data of the IRS 48 transition disk, revealing a wealth of molecular complexity.

\begin{itemize}
    \item We report the first detections of dimethy ether (\ce{CH_{3}OCH_{3}}), nitric oxide (NO), and a tentative detection of methyl formate (\ce{CH_{3}OCHO}) in a protoplanetary disk.
    \item We report an additional detection of a \ce{SO_2} transition in the disk with an upper energy level of $\approx$250~K. 
    \item The emissions of the detected species show a direct link with the asymmetric dust trap in the southern region of the disk, further suggesting that molecular complexity in this disk is due to ice sublimation.    
    \item The abundance ratios of \ce{CH_{3}OCH_{3}} and \ce{CH_{3}OCHO} compared with \ce{CH_3OH} are high relative to other environments. This either means that these molecules are enhanced relative to \ce{CH_3OH} in this disk or that the \ce{CH_3OH} column density we derive is underestimated. The latter situation could be due to the lines being optically thick but beam diluted. A higher \ce{CH_3OH} column density would mean a COM-emitting area smaller than the assumed area, the $5\sigma$ extent of the millimetre dust trap. With further high-angular-resolution observations, we will be able to determine whether or not the emitting area is truly just the thin inner edge of the dust trap. 
    \item The detection of \ce{CH_{3}OCH_{3}} and \ce{CH_{3}OCHO}  in such a warm disk and the agreement in the \ce{CH_{3}OCH_{3}}/\ce{CH_{3}OCHO} column density ratio with other environments strengthens the case for an origin inherited from the cold cloud phase, but the abundances with respect to \ce{CH_3OH} may be enhanced because of UV irradiation. 
    
\end{itemize}

Hopefully future observations of the IRS~48 icy dust trap will allow for the detection of other COMs and more robust constraints on the column density and excitation conditions. 
This work is an important puzzle piece in tracing the full interstellar journey of COMs across the different evolutionary stages of star, disk, and planet formation.

\begin{acknowledgements}

Astrochemistry in Leiden is supported by the Netherlands Research School for Astronomy (NOVA). 
ALMA is a partnership of ESO (representing its member states), NSF (USA) and NINS (Japan), together with NRC (Canada) and NSC and ASIAA (Taiwan) and KASI (Republic of Korea), in cooperation with the Republic of Chile. The Joint ALMA Observatory is operated by ESO, AUI/ NRAO and NAOJ. This paper makes use of the following ALMA data: 2013.1.00100.S, 2017.1.00834.S.

\end{acknowledgements}

\bibliographystyle{aa}
\bibliography{main}

\begin{appendix} 

\onecolumn

\section{Table of molecular lines detected}

\begin{center}
\begin{table*}[h!]
\caption{Properties of the molecular lines analysed in this work.}
\small
\centering
\begin{tabular}{clccccccc}
\hline \hline
Molecule & Transition & Frequency & $\mathrm{log_{10}(\textit{A}_{ij})}$ & $\mathrm{\textit{E}_{u}}$   & $\mathrm{\textit{g}_u}$ & Reference \\  
         &            &  (GHz)    & ($\mathrm{s^{-1}}$) & (K)              &  & & \\ 
         
\hline 
\ce{CH_{3}OH}  &  \ce{14_{ 1,13}} - \ce{14 _{0,14}}   & 349.106997  &  -3.356  & 260.2 & 116 &  \citet{van2021major} \\

\ce{CH_{3}OH^{[1]}}    &   \ce{4_{ 0, 4}} - \ce{3_{1, 3}}     & 350.687662 & -4.0619  & 36.3  & 36 & \citet{van2021major}\\

\ce{CH_{3}OH}   &   \ce{1_{1, 1}} - \ce{0_{0, 0}}     &  350.905100 & -3.4795  & 16.8 & 12 & \citet{van2021major}\\

\ce{CH_{3}OH^{[2]}}  &  \ce{9_{-5,4}} - \ce{9_{-4,6}},   & 351.236479  & -4.43803 &   240.5 & 76.0 & This work \\

\ce{CH_{3}OH}  &   \ce{11_{ 0,11}} - \ce{10_{1, 9}}  & 360.848946 & -3.9183  & 166.0 & 92 & \citet{van2021major}\\

\ce{CH_{3}OH^{[2]}}  &   \ce{3_{1,2}} - \ce{4_{2,2}}   & 361.236506 & -3.57681 & 339.2  & 28.0 &  This work\\

\ce{CH_{3}OH}  &  \ce{8_{1,7,1}} - \ce{7_{2,6,1}}   &  361.852195        & -4.1125       & 104.6 & 68&  \citet{van2021major} \\

\ce{CH_{3}OH}  &  \ce{16_{1,15}} - \ce{16_{ 0,16}}    &  363.440442  &  -3.3179 & 332.6 &132 & This work \\

\ce{CH_{3}OH}  &   \ce{7_{2,6}} - \ce{6_{1,5}}   &      363.739868 & -3.7677 & 87.3  & 60 &  \citet{van2021major}\\

\ce{CH_{3}OCH_{3}^{[3]}}  &   \ce{20_{1,20}} - \ce{19_{0,19}}    &  361.863445          &-3.4051       & 184.5 & 164 &This work \\

\ce{CH_{3}OCH_{3}^{[3]}}  &     \ce{20_{1,20}} - \ce{19_{0,19}} AE + EA   &  361.863446   &      -3.4051 & 184.5 & 246 &This work  \\

\ce{CH_{3}OCH_{3}^{[3]}}  &   \ce{20_{1,20}} - \ce{19_{0,19}} EE  &  361.863547          &       -3.4051        & 184.5 & 656  &  This work \\

\ce{CH_{3}OCH_{3}^{[3]}}  &  \ce{20_{1,20}} - \ce{19_{0,19}} AA   &  361.863648          &      -3.405  & 184.5 & 410 & This work \\

\ce{CH_{3}OCH_{3}^{[3]}}  &   \ce{20_{1,20}} - \ce{19_{0,19}} AE   &  361.863693         &       -3.2054 & 184.5 & 123 & This work \\

\ce{CH_{3}OCH_{3}^{[3]}}  &  \ce{20_{1,20}} - \ce{19_{0,19}} EE   &  361.863782          &      -3.2055 & 184.5  & 328 & This work \\

\ce{CH_{3}OCH_{3}^{[3]}}  &    \ce{20_{1,20}} - \ce{19_{0,19}} AA   & 361.863871         &       -3.2055 & 184.5 & 205 & This work \\

\ce{CH_{3}OCH_{3}}  &   \ce{11_{3, 8}} - \ce{10_{2, 9}}  AE    &  361.871157          &              -3.6652 & 72.9 & 69 & This work \\

\ce{CH_{3}OCHO^{[4]}}  & \ce{32_{3, 30}} - \ce{31_{3, 29}} E  &   363.482855          &      -3.16595                & 303.9 & 65 &This work \\

\ce{CH_{3}OCHO^{[4]}}  & \ce{32_{3, 30}} - \ce{31_{2, 29}} E  &   363.482881          &      -3.14442        &       303.9 & 130 & This work \\

\ce{CH_{3}OCHO^{[4]}}   & \ce{32_{2, 30}} - \ce{31_{2, 29}} E  &  363.485165          &      -3.16594        & 303.9  & 65 & This work \\

\ce{CH_{3}OCHO^{[4]}}  &   \ce{32_{2, 30}} - \ce{31_{2, 29}} E &  363.485187          &      -3.14442        & 303.9  & 130 &  This work\\

\ce{CH_{3}OCHO^{[4]}}   &  \ce{32_{2, 30}} - \ce{31_{3, 29}} A   &  363.487618          &      -3.89252        & 303.9   & 65 &  This work\\

\ce{CH_{3}OCHO^{[4]}} &  \ce{32_{2, 30}} - \ce{31_{3, 29}} A  &   363.487661          &              -4.02862 & 303.9  & 130 &  This work\\

\ce{CH_{3}OCHO^{[4]}} &   \ce{32_{3, 30}} - \ce{31_{2, 29}} A  &   363.488265    & -3.89249      & 303.9  & 65 &  This work\\

\ce{CH_{3}OCHO^{[4]}} &  \ce{32_{3, 30}} - \ce{31_{2, 29}} A  &         363.488271          &      -4.02872  & 303.9  & 130 & This work \\

\ce{CH_{3}OCHO^{[4]}}  &  \ce{32_{3, 30}} - \ce{31_{3, 29}} E &  363.490703          &              -3.16593        & 303.9  & 65 &  This work\\

\ce{CH_{3}OCHO^{[4]}} &   \ce{32_{3, 30}} - \ce{31_{3, 29}} E &  363.490764          &              -3.14442 & 303.9  &  130 &  This work\\

\ce{NO^{[1]} }  &    \ce{4_{1}} - \ce{4_{3}}     &  350.689494   &      -5.2662 & 36.2 & 10.0 &  This work \\

\ce{NO^{[1]} }  &    \ce{4_{1}} - \ce{4_{3}}     &  350.690766   &      -5.3032 & 36.2 & 8.0 &  This work\\

\ce{NO^{[5]} }  &    \ce{4_{1}} - \ce{4_{3}}     &  351.043524   &      -5.2649 & 36.2 & 10.0 &  This work \\

\ce{NO^{[5]} }  &    \ce{4_{1}} - \ce{4_{4}}     &  351.051705   &      -5.3019 & 36.2 & 8.0 & This work\\

\ce{NO^{[5]} }  &    \ce{4_{1}} - \ce{4_{3}}     &  351.051705   &      -5.316  & 36.2 & 6.0 & This work \\

\ce{SO_{2} }  &     \ce{21_{4,18}} - \ce{21_{3,19}}       &  363.159262          &       -3.397  &  252.1 & 43.0 &  This work \\

\hline
\end{tabular}
\label{Tab:1}
\begin{tablenotes}\footnotesize
\item{The line frequencies, Einstein A coefficients, upper energy levels ($E_{\rm{up}}$), and degeneracies ($\mathrm{g_u}$) are taken from the Cologne Database for Molecular Spectroscopy (CDMS)\,\citep{muller2001cologne, muller2005cologne,pickett1998submillimeter}. 
\ce{^{[1]}}Blended methanol and nitric oxide lines. \ce{^{[2]}} Weak methanol lines fitted at a different column density.\ce{^{[3]}}Blended dimethyl ether lines. \ce{^{[4]}}Blended methyl formate lines. \ce{^{[5]}}Blended nitric oxide lines.} 
\end{tablenotes}
\end{table*}  
\end{center}

\newpage

\section{Spectra}

\begin{figure*}[h!]
    \centering
    \includegraphics[trim={0 0 0 1cm},clip, width=\hsize]{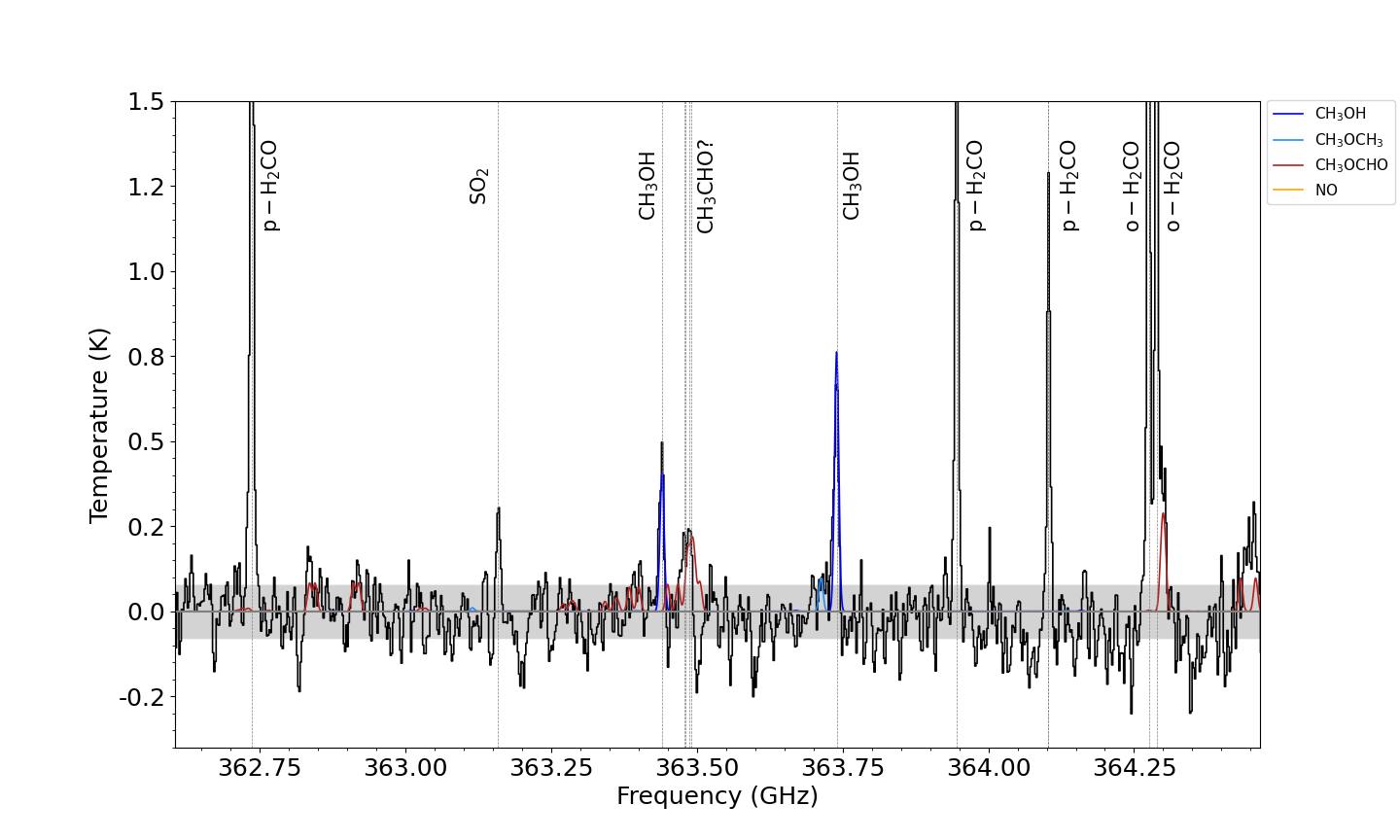}
    \caption{Stacked spectra with CASSIS model fits with $T_{\rm{ex}}$ = 100~K for all species aside from NO which is modelled at 40~K. The grey region shows the +/- 1$\sigma$ error. The vertical dashed lines denote the rest frequency of the lines. The \ce{CH_3OH} model is with a column density of 5$\times10^{14}$~cm$^{-2}$. In Figure~2 we show how a higher column density better fits weaker \ce{CH_3OH} lines covered in the observations.}
    \label{fig:spw0}
\end{figure*}

\begin{figure*}
    \centering
    \includegraphics[trim={0 0 0 1cm},clip, width=\hsize]{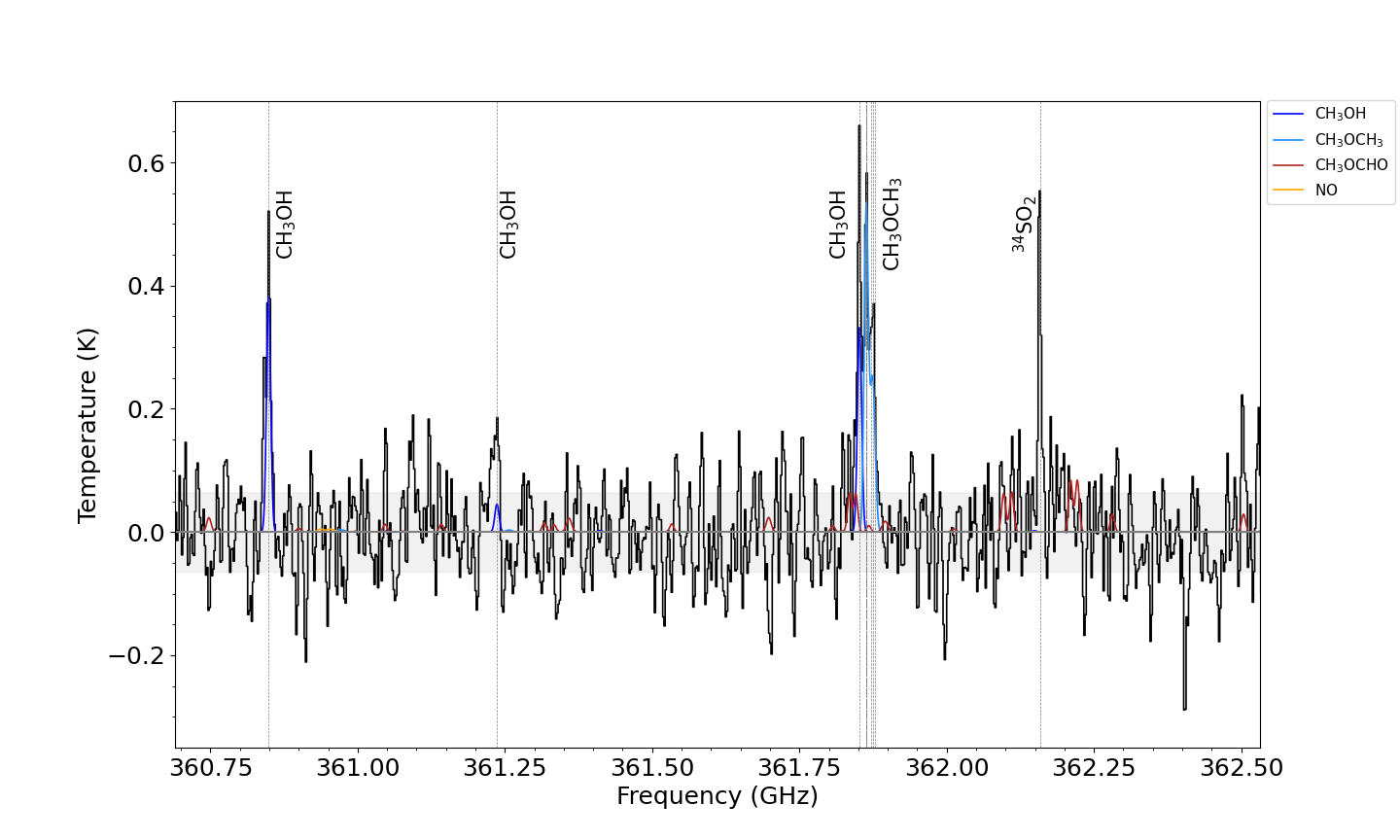}
    \caption{Same as Figure~\ref{fig:spw0}}   
    \label{fig:spw1}
\end{figure*}

\begin{figure*}[h!]
    \centering
    \includegraphics[trim={0 0 0 1cm},clip, width=\hsize]{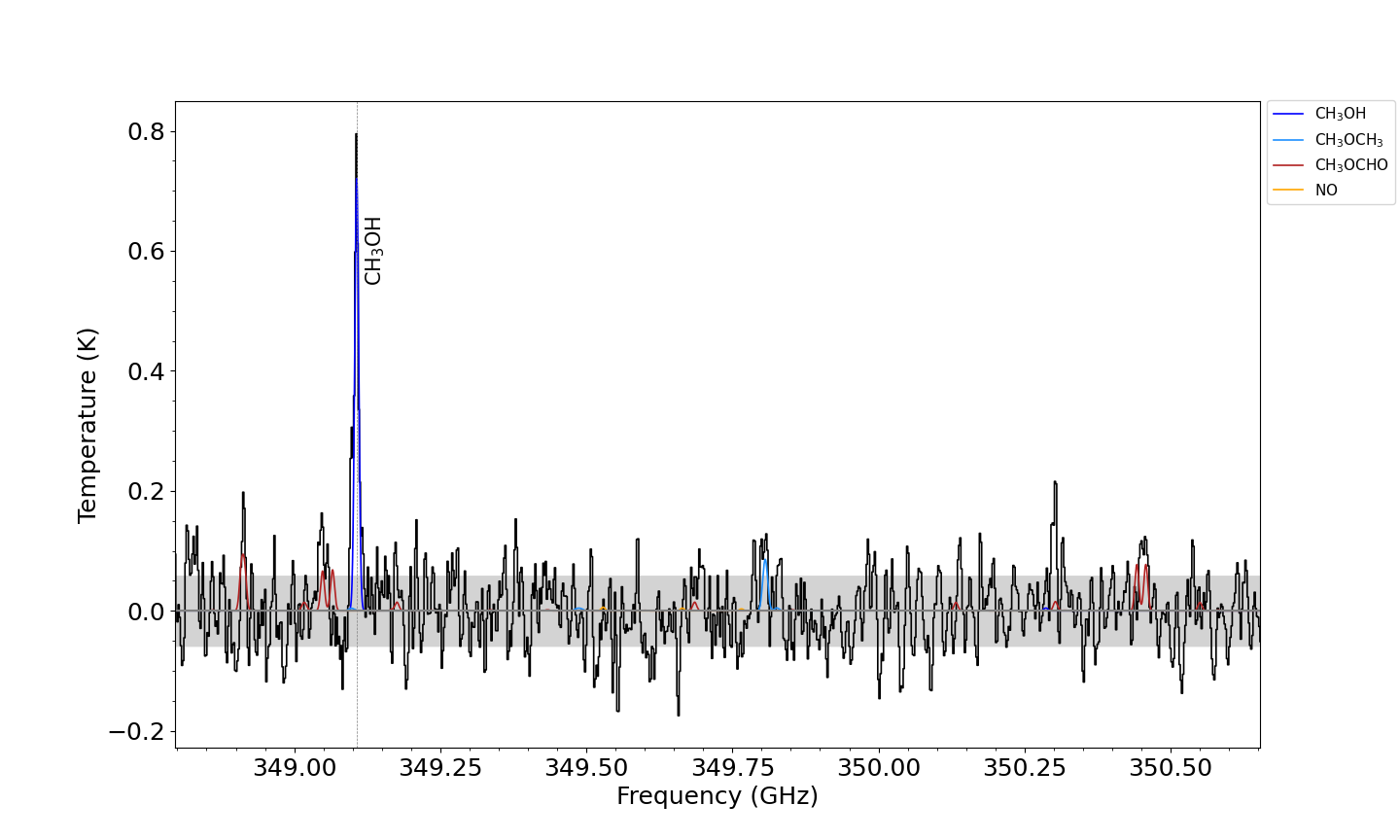}
    \caption{Same as Figure~\ref{fig:spw0}}   
  \label{fig:spw2}
\end{figure*}

\begin{figure*}[h!]
    \centering
    \includegraphics[trim={0 0 0 1cm},clip, width=\hsize]{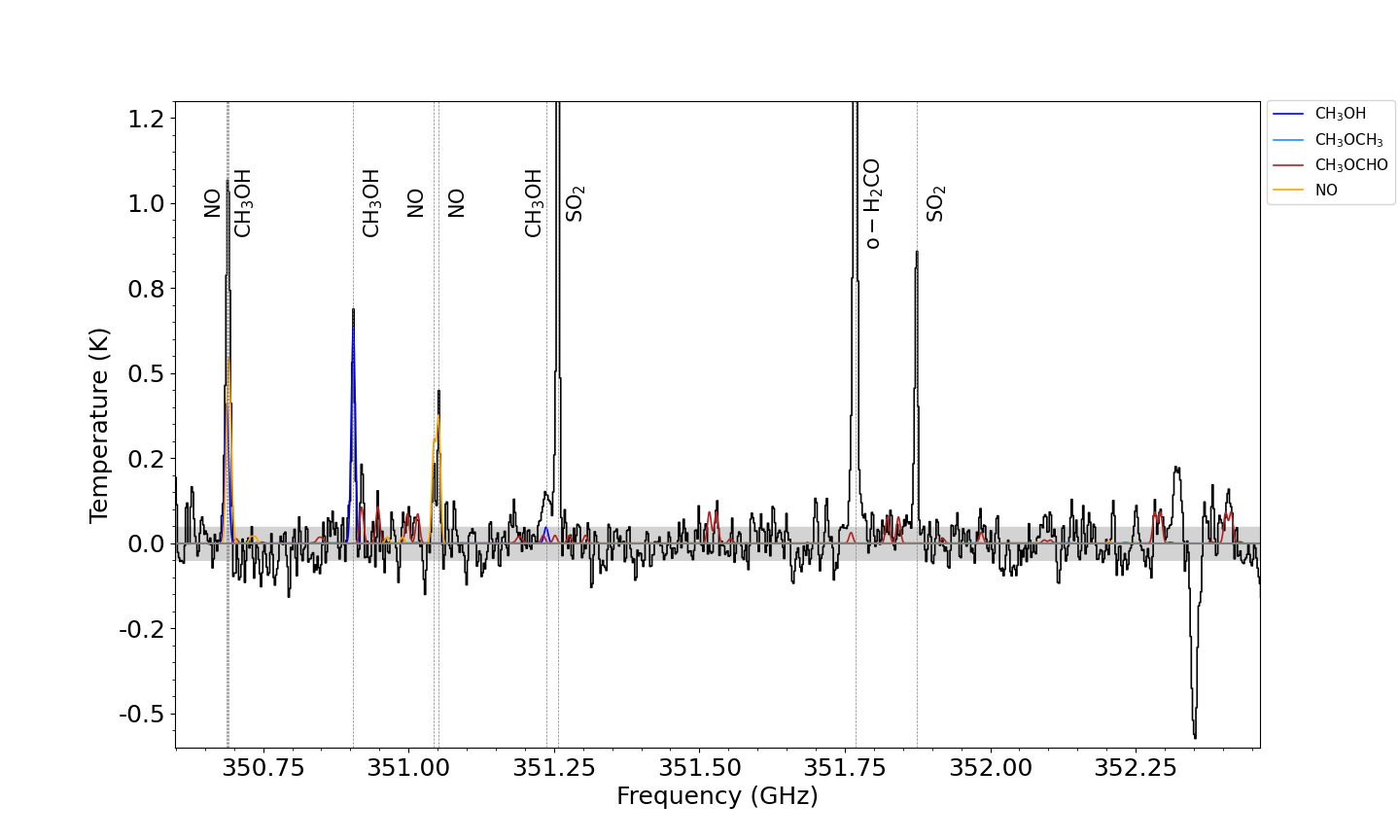}
    \caption{Same as Figure~\ref{fig:spw0}}   
  \label{fig:spw3}
\end{figure*}

\newpage

\section{Channel maps}

\begin{figure*}[h!]
    \centering
     \includegraphics[width=\hsize]{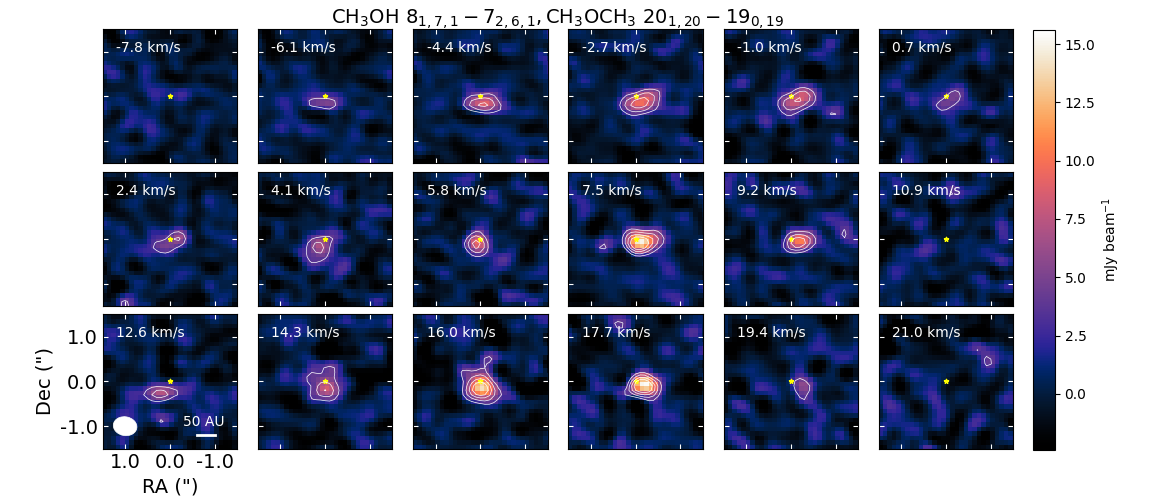}
     \caption{Channel maps of blended dimethyl ether and methanol lines. The first two rows show the two sets of dimethyl ether transitions while the bottom row shows emission coming from the methanol. The beam is shown in the bottom left corner and the scale bar is shown in the bottom right corner. Contours show the [3,5,7,9]$\times \sigma$ levels. } 
    \label{fig:channelether}
\end{figure*}

\begin{figure*}[h!]
    \centering
     \includegraphics[width=\hsize]{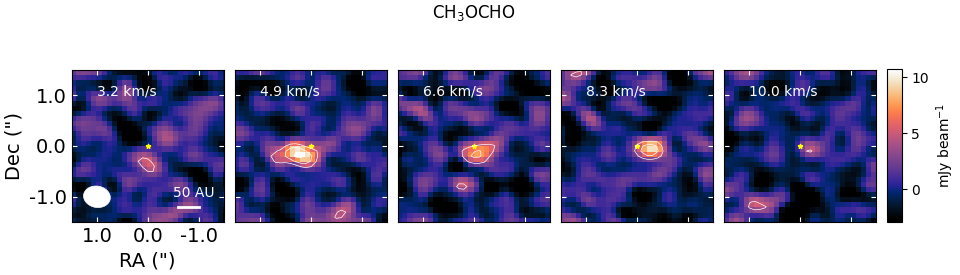}
    \caption{Channel maps of the methyl formate detection. The beam is shown in the bottom left corner and the scale bar is shown in the bottom right corner. Contours show the [3,5]$\times \sigma$ levels. } 
    \label{fig:channel maps}
\end{figure*}

\end{appendix}

\end{document}